\documentclass[11pt]{article}

\usepackage{fontspec}
\defaultfontfeatures{Mapping=tex-text}
\usepackage{xunicode}
\usepackage{xltxtra}

\usepackage[a4paper, margin=2cm]{geometry}

\usepackage[english]{babel}

\usepackage{amsmath, amssymb, bm}
\usepackage{slashed}

\usepackage{graphicx}
\usepackage{float}
\usepackage{subcaption}
\usepackage{booktabs}
\usepackage{siunitx}
\usepackage{multirow}

\usepackage{titlesec}
\titleformat{\section}
{\normalfont\Large\bfseries\centering}{\Roman{section}.}{1em}{}

\usepackage{cite}
\usepackage{indentfirst}
\usepackage{appendix}

\begin{document}

\begin{center}
	{\LARGE \textbf{Order-$v^2$ relativistic corrections to heavy-quark fragmentation into $P$-wave quarkonium states
	}} \\[12pt]

      \large 
      Sai Cui$^{(a)}$, Sheng-Juan Jiang$^{(a)}$, Guang-Zhi Xu$^{(a),*}$ and Kui-Yong Liu$^{(b,a),\dagger}$
      
      \vspace{6pt} 
      
      \normalsize 
      $^{(a)}$School of Physics, Liaoning University, Shenyang 110036, China\\
      $^{(b)}$School of Physics and Electronic Technology, Liaoning Normal University, Dalian 116029, China
      
      \vspace{6pt} 
      
      $*$Corresponding author. Email: xuguangzhi@lnu.edu.cn\\
      $\dagger$Corresponding author. Email: liukuiyong@lnu.edu.cn\\
	{\small \today}
\end{center}

	\begin{abstract}
		Within the framework of nonrelativistic QCD (NRQCD) factorization,
		and based on the Collins--Soper operator definition of fragmentation functions,
		we present a systematic calculation of the fragmentation functions for a heavy quark
		fragmenting into color-singlet $P$-wave quarkonium states.
		After reproducing and confirming the known leading-order results,
		we further compute the relativistic corrections up to order $\mathcal{O}(v^{2})$.
		Our analysis applies both to quarkonium systems composed of heavy quarks with the same flavor
		and to $B_c$-type mesons formed by heavy quarks of different flavors.
		Numerical results show that,
		for all color-singlet $P$-wave channels,
		the $\mathcal{O}(v^{2})$ relativistic corrections give sizable negative contributions
		over most of the momentum-fraction $z$ region.
        We further compute inclusive cross sections for $P$-wave quarkonium plus charmed hadrons in $e^+e^-$ annihilation via the single photon process up to $\mathcal{O}(v^{2})$ by applying our obtained fragmentation functions, and the resulting predictions are consistent with the full fixed-order results in the high-energy region.
	
	\end{abstract}
	
	\section{Introduction}
	
In high-energy collisions, the production of hadrons with large transverse
momentum is dominated by parton fragmentation.
In the context of heavy quarkonium production, the underlying production
mechanisms and factorization properties have been extensively investigated,
together with higher-order QCD and relativistic corrections in various channels
\cite{
	Brambilla:2010cs,
	Kang:2011mg,
	Gong:2008ft,
	Ma:2008gq,
	Zhang:2009ym,
	He:2009uf,
	Xu:2012am,
	Li:2013csa,
	Xu:2014zra
}.
According to the QCD factorization theorem,
the corresponding differential cross sections can be expressed
as convolutions of partonic production rates and fragmentation functions,
with nonperturbative hadronization effects absorbed into the latter
\cite{Collins:1981uw,Collins1989}.

	For heavy quarkonium systems, the small relative velocity $v$ of the heavy
	quark constituents introduces a clear hierarchy of energy scales.
	The separation between the short-distance scale $1/m_Q$
	and the bound-state formation scale $1/(m_Q v)$
	provides the theoretical foundation for nonrelativistic QCD (NRQCD)
	\cite{CaswellLepage1986}.
	Within this framework, quarkonium production processes can be factorized
	into short-distance coefficients(SDCs) and long-distance matrix elements(LDMEs),
	organized as double expansions in $\alpha_s$ and $v$
	\cite{BBL1995}.
	
	The fragmentation functions for heavy quarkonium have been extensively
	studied at leading order(LO) within the NRQCD factorization formalism.
	Results are available for both gluon and heavy-quark fragmentation
	into $S$-wave and $P$-wave quarkonium states,
	and have been further extended to higher orbital angular momentum channels,
	including $D$-wave states
	\cite{Braaten:1993rw,Braaten:1993mp,Braaten:1994kd,Yuan:1994hn,
		Ma:1995ci,Ma:1995vi,Cho:1994qp,Cheung:1995ir,Qiao:1997wb}.
	
	Beyond the LO, QCD corrections to fragmentation functions
	have been computed for a variety of channels.
	In particular, next-to-leading-order(NLO) QCD corrections
	are now available for several $S$-wave fragmentation processes
	and have been applied to both quarkonium and $B_c$ production
	\cite{Zhang:2018mlo,Zheng:2019gnb,Feng:2018ulg,
		Zheng:2021sdo,Feng:2021qjm,Zhang:2021ypo}.
	Related studies have also investigated the production of higher
	orbital excitations in hadron colliders
	\cite{Berezhnoy:2021wrc}.
	
Relativistic corrections constitute another essential ingredient
for improving the theoretical accuracy of fragmentation functions.
For $S$-wave states, relativistic corrections at
$\mathcal{O}(v^2)$ have been systematically studied,
and sizable effects have been observed in both quark-
and selected gluon-fragmentation channels
\cite{Bodwin:2003wh,Sang:2009zz,Gao:2016ihc}.
Higher-order relativistic corrections have been studied in detail, with complete
$\mathcal{O}(v^4)$ analyses carried out for gluon-fragmentation processes
\cite{Bodwin:2012xc} as well as for heavy-quark fragmentation channels
\cite{Cui:2025wjq}.
In contrast, relativistic corrections to heavy-quark fragmentation
into $P$-wave quarkonium remain largely unexplored.
Existing studies are limited to LO results,
and a complete $\mathcal{O}(v^2)$ analysis
has not yet been carried out for these channels
\cite{Yuan:1994hn,Ma:1995vi}.
	
	For systems composed of heavy quarks with unequal masses,
	such as the $B_c$ family,
	relativistic effects are expected to interplay with mass asymmetry,
	potentially leading to additional modifications of fragmentation dynamics.
Although fragmentation functions for $B_c$ mesons
have been studied at LO and at NLO
in $\alpha_s$
\cite{Zheng:2021sdo,Feng:2021qjm,Zheng:2019gnb},
a systematic treatment of relativistic corrections
for $P$-wave $B_c$ states is still missing.

In this work, we present a systematic analysis of heavy-quark fragmentation
into all color-singlet(CS) $P$-wave heavy hadrons within the NRQCD factorization
framework.
After reproducing and confirming the known LO results
for heavy-quark fragmentation into $P$-wave states
\cite{Yuan:1994hn,Ma:1995vi},
we derive the complete $\mathcal{O}(v^2)$ relativistic corrections
to the corresponding fragmentation functions.
Our study applies to both equal-mass quarkonium systems
and unequal-mass configurations, including the $B_c$ family.
These results close an important gap in the theoretical understanding
of $P$-wave fragmentation functions
and improve the precision of theoretical predictions
for $P$-wave heavy-hadron production
in the large transverse-momentum region.

    The remainder of this paper is organized as follows.
	In Sec.~II, we briefly review the gauge-invariant Collins--Soper definition of fragmentation functions.
	In Sec.~III, we introduce the NRQCD factorization framework relevant for heavy-quark fragmentation.
	In Sec.~IV, we present a unified calculation of heavy-quark fragmentation functions into CS $P$-wave heavy mesons in the axial gauge, with a formulation that applies to both equal-mass quarkonium systems and unequal-mass configurations, including the $B_c$ family.
	In Sec.~V, we extract the corresponding SDCs through matching to the NRQCD expansion.
	In Sec.~VI, we provide numerical results and discussions, with particular emphasis on the impact of the $\mathcal{O}(v^2)$ relativistic corrections for different $P$-wave channels.
	In Sec.~VII, we present a comparison between the fragmentation approximation and the full calculation for $P$-wave charmonium production.
	In Sec.~VIII, we summarize our main results.
	In Appendix~A, we present the expansion of the squared amplitude for $P$-wave relativistic corrections, using the equal-mass case as an explicit example.
	In Appendices~B and~C, we list the SDCs for all $P$-wave states in the equal-mass and unequal-mass cases, respectively.

	\section{Collins--Soper definition of fragmentation functions}
	\label{Collins--Soper}
	
	To compute the fragmentation functions for heavy-quark fragmentation
	into $P$-wave heavy quarkonium,
	we adopt the gauge-invariant operator definition
	proposed by Collins and Soper~\cite{Collins:1981uw}.
	The calculation is carried out in the light-cone coordinate system,
	which allows a transparent separation of longitudinal and transverse
	degrees of freedom and is therefore convenient for analyzing
	the kinematic structure of the fragmentation process.
	
	In light-cone coordinates,
	an arbitrary four-vector $p^\mu$ is written as
	\begin{align}
		&p^\mu= (p^+, p^-, \mathbf{p_T}) , \label{eq:p_def} \\
		&	p^+ = (p^0 + p^3)/\sqrt{2}, \label{eq:p_plus} \\
		&p^- = (p^0 - p^3)/\sqrt{2}. \label{eq:p_minus}
	\end{align}
	The scalar product of two four-momenta $p$ and $q$ is expressed as
	\begin{equation}
		p \cdot q = p^+ q^- + p^- q^+ -\mathbf{p_T} \cdot\mathbf{q_T} .
	\end{equation}
	We introduce a light-cone vector
	\begin{equation}
	\hat	n^{\mu} = (0,\,1,\,\mathbf{0_T}),
	\end{equation}
	with $\hat n^2 = 0$.
	Based on this, the gauge-invariant fragmentation function for a heavy quark fragmenting into a quarkonium state $H$ is defined by Collins and Soper as \cite{Collins:1981uw}
\begin{equation}
	\begin{aligned}
		D_{H/Q}(z)
		&= \frac{z^{\,d-3}}{4\pi}
		\int dx^{-}\, e^{- i P^{+} x^{-}/z}\,
		\frac{1}{3}\mathrm{Tr}_{\text{color}}
		\,\frac{1}{2}\mathrm{Tr}_{\text{Dirac}}
		\\[4pt]
		&\times
		\Biggl\{
		\hat n\!\cdot\!\gamma\;
		\langle 0|
		Q(0)\,
		\bar {\text{P}} \exp\!\left[
		-i g_{s}\!\int_{0}^{\infty}\! d\lambda\;
	\hat	n\!\cdot\! A^{T}(\lambda \hat n^\mu)
		\right]
		\\[4pt]
		&\times
		a_{H}^{\dagger}(P^{+},\mathbf{0_{T}})\,
		a_{H}(P^{+},\mathbf{0_{T}})\,
		\text{P}\exp\!\left[
		i g_{s}\!\int_{x^{-}}^{\infty}\! d\lambda\;
	\hat	n\!\cdot\! A^{T}(\lambda \hat n^\mu)
		\right]
		\\[4pt]
		&\times
		\bar Q(0, x^{-}, \mathbf{0_{T}})
		|0\rangle
		\Biggr\}.
		\qquad 
	\end{aligned}
\end{equation}

In this definition, the longitudinal momentum fraction is given by
$
	z = \frac{P^{+}}{k^{+}},
$
where $k^{+}$ denotes the light-cone ``$+$'' component of the fragmenting heavy-quark momentum, and $P^{+}$ is that of the produced hadron $H$.
The calculation is performed in $d = 4 - 2\epsilon$ space-time dimensions using dimensional regularization.
Since no ultraviolet or infrared divergences appear in the present analysis, we set $d = 4$ in the final expressions.
The path-ordered exponential, $\mathrm{P}\exp\{\cdots\}$, corresponds to a Wilson line and ensures the gauge invariance of the nonlocal operator.
The operator $a_H^\dagger$ creates the hadronic state $H$, and $A_\mu$ denotes the gluon field.

	\section{NRQCD factorization formalism}
	\label{NRQCD}
	
	In the large transverse-momentum region,
	the production of bound states composed of a heavy quark and a heavy antiquark
	is dominated by the fragmentation mechanism.
	According to the QCD factorization theorem,
	the corresponding differential cross section
	can be expressed as a convolution of the partonic production cross section
	and the fragmentation function,
	a structure that directly follows from QCD factorization
	\cite{Collins:1981uw,Collins1989}.

	When the final state is a bound system formed by a heavy quark and a heavy antiquark,
	either of equal or unequal flavors,
	the presence of the heavy-quark mass scale,
	which is much larger than $\Lambda_{\mathrm{QCD}}$,
	introduces several well-separated energy scales
	in the fragmentation process.
	These include the heavy-quark mass scale $m$,
	the typical relative-momentum scale $m v$,
	and the nonrelativistic kinetic-energy scale $m v^{2}$.
	In the nonrelativistic limit $v \ll 1$,
	these scales satisfy the hierarchy
	\begin{equation}
		m v^{2} \ll m v \ll m .
	\end{equation}
	Exploiting this clear separation of scales,
	the NRQCD effective field theory
	integrates out the high-energy degrees of freedom at the scale $m$,
	thereby systematically separating
	the short-distance production of the heavy quark--antiquark pair
	from the long-distance nonperturbative dynamics
	associated with bound-state formation
	\cite{CaswellLepage1986,BBL1995}.
	Within this framework,
	all nonperturbative effects related to hadronization
	are absorbed into NRQCD LDMEs,
	allowing fragmentation functions
	to be further factorized within the NRQCD formalism.
	
	Accordingly,
	the fragmentation function for a heavy quark $Q$
	fragmenting into a hadronic state $H$
	can be written as a sum of products
	of SDCs
	and the corresponding NRQCD LDMEs
	\begin{equation}
		D_{H/Q}(z)
		=
		\sum_{n}
		\left[
		F^{n}(z)\,
		\langle 0 | \mathcal{O}^{H}(n) | 0 \rangle
		+
		G^{n}(z)\,
		\langle 0 | \mathcal{P}^{H}(n) | 0 \rangle
		\right]
		+
		\mathcal{O}(v^{4}) \, ,
	\end{equation}
	where $F^{n}(z)$ and $G^{n}(z)$
	are SDCs
	that can be calculated perturbatively in QCD.
	The matrix elements
	$\langle 0|\mathcal{O}^{H}(n)|0\rangle$
	and
	$\langle 0|\mathcal{P}^{H}(n)|0\rangle$
	denote the NRQCD LDMEs
	corresponding to the intermediate state $n$.
	
	In this work,
	we focus on the production of $P$-wave heavy-quark bound states.
	Under the leading Fock-state approximation,
	only CS NRQCD production operators
	with the same quantum numbers as the hadronic state $H$
	are retained.
	For $P$-wave states,
	the dominant CS channels include
	$^{1}P_{1}^{[1]}$
	and
	$^{3}P_{J}^{[1]}$ with $J=0,1,2$.
	Within the NRQCD factorization framework,
	relativistic corrections are described by higher-dimensional NRQCD operators
	that carry the same quantum numbers
	and are organized as an expansion in powers of
	the relative velocity $v$ of the heavy quarks.
	The corresponding SDCs
	share the same dynamical structure as those at LO
	and can be determined through matching calculations
	performed for the fragmentation of a free heavy quark--antiquark pair.
	This procedure allows one to consistently obtain
	the $\mathcal{O}(v^{2})$ and higher-order relativistic corrections
	to the production of $P$-wave bound states.
	
	Based on the above discussion,
	the NRQCD production operators
	corresponding to hadronic states with quantum numbers
	$^{1}P_{1}^{[1]}$ and $^{3}P_{J}^{[1]}$
	are listed below.
	\begin{align}
		\mathcal{O}^H(^1P_1)
		&=\frac{1}{N_c}\,
		\chi^\dagger\!\left(-\frac{i\overleftrightarrow{\mathbf{D}}}{2}\right)\!\psi
		\sum_X |H+X\rangle\langle H+X|
		\psi^\dagger\!\left(-\frac{i\overleftrightarrow{\mathbf{D}}}{2}\right)\!\chi,
		\\[6pt]
		\mathcal{P}^H(^1P_1)
		&=\frac{1}{2N_c}
		\left[
		\chi^\dagger\!\left(-\frac{i\overleftrightarrow{\mathbf{D}}}{2}\right)
		\left(\frac{i\overrightarrow{\mathbf{D}}}{2}\right)^2\!
		\psi\sum_X |H+X\rangle\langle H+X|
		\psi^\dagger\!\left(-\frac{i\overleftrightarrow{\mathbf{D}}}{2}\right)\chi
		+{\rm H.c.}
		\right],
		\\[8pt]
		\mathcal{O}^H(^3P_0)
		&=\frac{1}{3N_c}\,
		\chi^\dagger\!\left(-\frac{i\overleftrightarrow{\mathbf{D}}\cdot\sigma}{2}\right)\!\psi
		\sum_X |H+X\rangle\langle H+X|
		\psi^\dagger\!\left(-\frac{i\overleftrightarrow{\mathbf{D}}\cdot\sigma}{2}\right)\!\chi,
		\\[6pt]
		\mathcal{P}^H(^3P_0)
		&=\frac{1}{6N_c}
		\left[
		\chi^\dagger\!\left(-\frac{i\overleftrightarrow{\mathbf{D}}\cdot\sigma}{2}\right)
		\left(\frac{i\overrightarrow{\mathbf{D}}}{2}\right)^2\psi
		\sum_X |H+X\rangle\langle H+X|
		\psi^\dagger\!\left(-\frac{i\overleftrightarrow{\mathbf{D}}\cdot\sigma}{2}\right)\chi
		+{\rm H.c.}
		\right],
		\\[8pt]
		\mathcal{O}^H(^3P_1)
		&=\frac{1}{2N_c}\,
		\chi^\dagger\!\left(-\frac{i\overleftrightarrow{\mathbf{D}}\times\sigma}{2}\right)\!\psi
		\sum_X |H+X\rangle\langle H+X|
		\psi^\dagger\!\left(-\frac{i\overleftrightarrow{\mathbf{D}}\times\sigma}{2}\right)\!\chi,
		\\[6pt]
		\mathcal{P}^H(^3P_1)
		&=\frac{1}{4N_c}
		\left[
		\chi^\dagger\!\left(-\frac{i\overleftrightarrow{\mathbf{D}}\times\sigma}{2}\right)
		\left(\frac{i\overrightarrow{\mathbf{D}}}{2}\right)^2\psi
		\sum_X |H+X\rangle\langle H+X|
		\psi^\dagger\!\left(-\frac{i\overleftrightarrow{\mathbf{D}}\times\sigma}{2}\right)\chi
		+{\rm H.c.}
		\right],
		\\[8pt]
		\mathcal{O}^H(^3P_2)
		&=\frac{1}{N_c}\,
		\chi^\dagger\!\left(-\frac{i\overleftrightarrow{D}^{(i}\sigma^{j)}}{2}\right)\!\psi
		\sum_X |H+X\rangle\langle H+X|
		\psi^\dagger\!\left(-\frac{i\overleftrightarrow{D}^{(i}\sigma^{j)}}{2}\right)\!\chi,
		\\[6pt]
		\mathcal{P}^H(^3P_2)
		&=\frac{1}{2N_c}
		\left[
		\chi^\dagger\!\left(-\frac{i\overleftrightarrow{D}^{(i}\sigma^{j)}}{2}\right)
		\left(\frac{i\overrightarrow{\mathbf{D}}}{2}\right)^2\psi
		\sum_X |H+X\rangle\langle H+X|
		\psi^\dagger\!\left(-\frac{i\overleftrightarrow{D}^{(i}\sigma^{j)}}{2}\right)\chi
		+{\rm H.c.}
		\right].
	\end{align}
	Here, $H$ denotes the heavy-quarkonium hadronic state in the final state,
	and $X$ represents the additional particles produced in association with $H$.
	The sum $\sum_X |H+X\rangle\langle H+X|$ corresponds to the completeness relation
	projecting onto the subspace of final states containing the hadron $H$.
	The number of colors is fixed to $N_c = 3$.
	The fields $\psi$ and $\chi$ denote the two-component Pauli spinor fields
	of the heavy quark and antiquark, respectively,
	and $\sigma^i\,(i=1,2,3)$ are the Pauli matrices.
	The operator $\mathbf{D}$ stands for the spatial components of
	the covariant derivative $D^\mu$.
	The combination
	\begin{equation}
		\overleftrightarrow{\mathbf{D}}
		\equiv
		\overrightarrow{\mathbf{D}}-\overleftarrow{\mathbf{D}}
	\end{equation}
	is introduced to encode the relative momentum
	of the heavy quark--antiquark pair in the NRQCD operators.
	
	The SDCs are determined through
	the standard matching procedure within the NRQCD factorization framework.
	In the matching, the physical hadronic state $H$
	is replaced by an on-shell free heavy quark--antiquark pair
	carrying the same overall quantum numbers,
	which may consist of either equal or unequal flavors.
	Under this replacement,
	the NRQCD LDMEs reduce to quantities
	that can be computed perturbatively,
	while the SDCs retain information
	solely from the hard scale
	and are insensitive to the long-distance bound-state dynamics.
	Consequently, by computing the fragmentation functions
	for the free heavy quark--antiquark pair
	and matching them term by term
	to the NRQCD operator expansion,
	the SDCs $F_n(z)$ and $G_n(z)$
	can be uniquely determined.

	\section{Fragmentation function calculation for $P$-wave heavy mesons}
	\label{axial-gauge}
	
	In this section, we present a unified formulation for the fragmentation
	of a heavy quark into CS $P$-wave heavy mesons
	within the Collins--Soper operator definition.
	The framework applies to both equal-mass quarkonium systems
	and unequal-mass states composed of two different heavy flavors,
	such as the $B_c$ family.
	
	The momenta of the heavy quark and antiquark forming the bound state
	are parameterized in terms of the total momentum $P$
	and the relative momentum $q$ as	\cite{Bodwin:2010fi}
	\begin{align}
		p_1 &= r P + q , \\
		p_2 &= (1-r) P - q ,
	\end{align}
	where the momentum fraction $r$ is fixed by the quark masses,
	\begin{equation}
		r = \frac{E_1}{E_1+E_2}.
	\end{equation}
	In the equal-mass limit $m_1=m_2$, one has $E_1=E_2$ and hence $r=1/2$.
	For a general unequal-mass system,
	the invariant mass of the $Q_1\bar Q_2$ pair reads
	\begin{equation}
		P^2 = M^2 = (E_1 + E_2)^2 ,
	\end{equation}
	with
	\begin{align}
		E_1 &= \sqrt{m_1^2 +  \mathbf{q}^{\,2}}, \\
		E_2 &= \sqrt{m_2^2 +  \mathbf{q}^{\,2}} .
	\end{align}
	Here, $M$ denotes the mass of the charmonium state. In the nonrelativistic limit $\boldsymbol{q}^{\,2} \ll m_{1,2}^2$, \(M =m_1+m_2\) and $P^2 \simeq (m_1+m_2)^2$.
	For the equal-mass case, $m_1=m_2=m$, $E_1=E_2=E$.
    For the invariant magnitude of $\mathbf q$, there exists a relation associated with the effective velocity $v$, 
	\begin{equation}
		\mathbf q = 2 \mu  v,
	\end{equation}
	where
	\begin{equation}
		\mu = \frac{m_1 m_2}{m_1 + m_2}
	\end{equation}
	is the reduced mass of the heavy quark--antiquark system.
	
	The spin-singlet and spin-triplet covariant projection operators
	for a general heavy-quark system are given by
	\begin{align}
		\Pi_{0} &=
		\frac{1}{4\sqrt{2N_c\,E_1E_2(E_1+m_1)(E_2+m_2)}}
		(\slashed{p}_2-m_2)\gamma^5
		\frac{\slashed{P}+E_1+E_2}{E_1+E_2}
		(\slashed{p}_1+m_1) ,
		\\
		\Pi_{1} &=
		\frac{1}{4\sqrt{2N_c\,E_1E_2(E_1+m_1)(E_2+m_2)}}
		(\slashed{p}_2-m_2)\gamma^\mu
		\frac{\slashed{P}+E_1+E_2}{E_1+E_2}
		(\slashed{p}_1+m_1) ,
	\end{align}
	where $N_c=3$ and Dirac spinors are normalized as $u^\dagger u=v^\dagger v=1$.
	In the equal-mass case,
	these expressions reduce to the standard projection operators
	commonly used for quarkonium states.

\begin{figure}[tbp]
		\centering %
		\includegraphics[width=.45\textwidth]{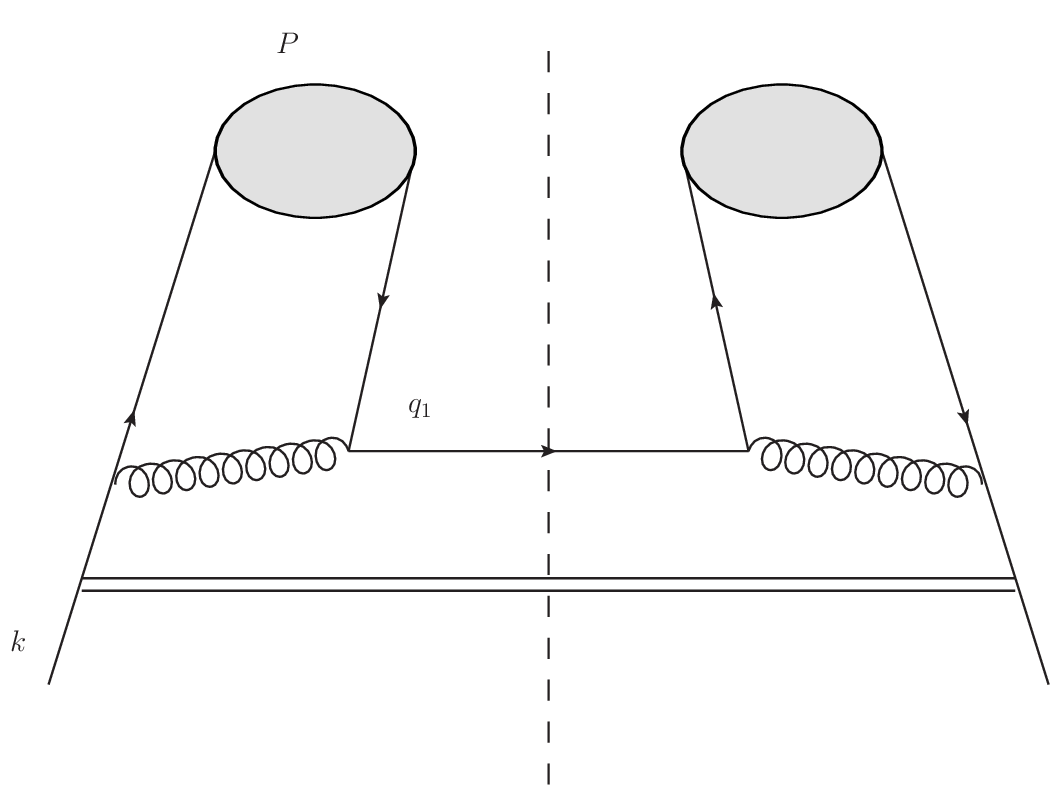}
		\caption{\label{fig:0} Feynman diagrams for heavy-quark fragmentation into heavy quarkonium at leading-order in $\alpha_s$, where the shaded blob denotes the heavy quarkonium state and the double line represents the Wilson line.
			}
	\end{figure}

	The fragmentation function for a heavy quark $Q_1$
	fragmenting into a pair of $Q_1\bar Q_2$ with quantum numbers
	${}^{2S+1}P_J$ can be written as
	\begin{align}
		D^{(^{2S+1}P_J)}_{(Q_1\bar Q_2)/Q_1}(z)
		&=
		2(E_1+E_2)\frac{z}
		{24\pi}
		\int
		\frac{dq_1^+\, d^2 \mathbf{q_{1T}}}
		{(2\pi)^3\,2q_1^+}
		\,
		2\pi\,\delta(k^+-P^+-q_1^+)\,
		F_c\,
		\mathcal A_{Q_1\to(Q_1\bar Q_2)[{}^{2S+1}P_J]} ,
	\end{align}
	where $q_1$ denotes the momentum of the recoiling final-state parton,
	and the color factor is
	\begin{equation}
		F_c=\mathrm{Tr}(T^aT^aT^bT^b)=\frac{16}{3}.
	\end{equation}	
	The factor $2(E_1+E_2)$ originates from the relativistic normalization of the NRQCD heavy-quark pair state.
	The squared amplitude  reads
	\begin{equation}
		\mathcal A_{Q_1\to(Q_1\bar Q_2)[{}^{2S+1}P_J]}
		=
		\mathrm{Tr}\!\left[
		\slashed{\hat n}\,
		\mathcal M_{P\text{-wave}}
		(\slashed q_1+m_2)\,
		\mathcal M_{P\text{-wave}}^*
		\right].
		\label{AP}
	\end{equation}
    Working in the axial gauge $\hat n \cdot A = 0$, 
	the Wilson line in the Collins--Soper definition vanishes, 
	and the fragmentation function receives contributions only 
	from diagrams involving a single gluon exchange, 
	as shown in Figure~\ref{fig:0}.
    The amputated amplitude is given by
	\begin{align}
		\mathcal M_{P\text{-wave}}
		&=
		D_{\mu\nu}(l)\,
		\frac{\slashed P+\slashed q_1+m_1}{(P+q_1)^2-m_1^2}\,
		\gamma^\mu\,
		\Pi_{0(1)}\,
		\gamma^\nu ,
	\end{align}
	where $D_{\mu\nu}(l)$ is the gluon propagator in the axial gauge,
	with the gluon momentum
	$	l = p_2 + q_1$ .
	The gluon propagator in the axial gauge takes the form
	\begin{equation}
		D_{\mu\nu}(l)
		=
		\frac{i}{l^2}
		\left(
		-g_{\mu\nu}
		+
		\frac{l_\mu \hat n_\nu + l_\nu \hat n_\mu}{l \!\cdot\! \hat n}
		\right).
	\end{equation}
	
    With this velocity counting, the $P$-wave amplitude can be systematically
	expanded in powers of $\mathbf q^{\,2}$ (or equivalently $v^2$) as
	\begin{equation}
		\mathcal M_{P\text{-wave}}
		=
		\mathcal M_{P1}
		+
		\mathbf q^{\,2}\,\mathcal M_{P3}
		+
		\mathcal O(\mathbf q^4).
		\label{MP}
	\end{equation}
	where $\mathcal M_{P1}$ and $\mathcal M_{P3}$ denote the leading and
	$\mathcal O(v^2)$ contributions, respectively.
	\begin{equation}
		\begin{aligned}
			&\mathcal M_{P1} = \epsilon^{\alpha}(L_z)\frac{\partial\mathcal M}{\partial q^{\alpha}} \Big|_{q \to 0}, \\
			&\mathcal M_{P3} =- \frac{1}{5\times 3!}\epsilon^{\alpha}(L_z)I_\alpha^{ \beta \gamma \delta}\frac{\partial^3\mathcal{M}}{\partial q^\beta\,\partial q^\gamma\,\partial q^\delta}\Big|_{q=0}
		\end{aligned}
	\end{equation}
    where
	\begin{equation}
		I^{\alpha\beta}
		=
		-g^{\alpha\beta}
		+
		\frac{P^{\alpha}P^{\beta}}{M^{2}}
	\end{equation}
	\begin{equation}
		I_\alpha^{\beta\gamma\delta}
		=
		I_\alpha^{\beta}I^{\gamma\delta}
		+
		I_\alpha^{\gamma}I^{\beta\delta}
		+
		I_\alpha^{\delta}I^{\beta\gamma}.
	\end{equation}
	We perform the expansion details at the squared amplitude level in Appendix A.

	\section{Matching of short-distance coefficients}
	\label{short-distance}
	
	To extract the SDCs for heavy-quark fragmentation
	into $P$-wave meson states,
	the perturbative fragmentation functions obtained in the previous section
	are matched term by term to the NRQCD factorization formula.
	With standard relativistic normalization,
	the NRQCD matrix elements for a heavy quark--antiquark pair
	with quantum numbers ${}^{2S+1}P_J$
	satisfy the normalization conditions
	\begin{align}
		\langle \mathcal{O}^{Q_1\bar Q_2}({}^1P_1)\rangle &= 2(d-1)N_c \, \mathbf{q}^2, \\
		\langle \mathcal{O}^{Q_1\bar Q_2}({}^3P_0)\rangle &= 2N_c\, \mathbf{q}^2, \\
		\langle \mathcal{O}^{Q_1\bar Q_2}({}^3P_1)\rangle &=2(d-1)N_c\, \mathbf{q}^2, \\
		\langle \mathcal{O}^{Q_1\bar Q_2}({}^3P_2)\rangle &=  (d+1) (d-2)N_c\, \mathbf{q}^2,
		\label{eq:Pwave_freeO}
	\end{align}
	and
	\begin{equation}
		\langle \mathcal{P}^{Q_1\bar Q_2}(n)\rangle
		=\mathbf{q}^2\,\langle \mathcal{O}^{Q_1\bar Q_2}(n)\rangle,
		\qquad n={}^1P_1,\,{}^3P_J~(J=0,1,2),
		\label{eq:Pwave_freeP}
	\end{equation}

	Substituting these matrix elements into the NRQCD factorization formula, the $P$-wave fragmentation function can be expressed as
	\begin{align}
		D_{Q_1\bar Q_2({}^1P_1)/Q_1}(z)
		&= 2(d-1)\,N_c\left(F^{\,{}^1P_1}(z)+\mathbf{q}^2\,G^{\,{}^1P_1}(z)\right),
		\label{eq:Pwave_frag_1P1}\\[4pt]
		D_{Q_1\bar Q_2({}^3P_0)/Q_1}(z)
		&= 2\,N_c\left(F^{\,{}^3P_0}(z)+\mathbf{q}^2\,G^{\,{}^3P_0}(z)\right),
		\label{eq:Pwave_frag_3P0}\\[4pt]
		D_{Q_1\bar Q_2({}^3P_1)/Q_1}(z)
		&=2(d-1)\,N_c\left(F^{\,{}^3P_1}(z)+\mathbf{q}^2\,G^{\,{}^3P_1}(z)\right),
		\label{eq:Pwave_frag_3P1}\\[4pt]
		D_{Q_1\bar Q_2({}^3P_2)/Q_1}(z)
		&=  (d+1) (d-2)\,N_c\left(F^{\,{}^3P_2}(z)+\mathbf{q}^2\,G^{\,{}^3P_2}(z)\right).
		\label{eq:Pwave_frag_3P2}
	\end{align}
	where $F^{n}(z)$ and $G^{n}(z)$ correspond to the LO and NLO SDCs in the $v$ expansion, respectively.
	By comparing the above expression term by term with the fragmentation function obtained from perturbative calculations in the previous section, the SDCs for all $P$-wave states with quantum numbers ${}^{2S+1}P_{J}$ can be determined.

	We present the SDCs for all $P$-wave states
	in both the equal-mass and unequal-mass cases.
For completeness, the explicit results for the equal-mass case are collected in
Appendix~B, while those for the unequal-mass case are presented in Appendix~C.
Our LO results are in agreement with those obtained by
	Yuan~\cite{Yuan:1994hn} and Ma~\cite{Ma:1995vi}.
	Going beyond the LO,
	we have systematically computed the $\mathcal{O}(v^2)$ relativistic corrections
	to heavy-quark fragmentation into $P$-wave states.
	For the equal-mass quarkonium systems,
	the $\mathcal{O}(v^2)$ relativistic corrections
	presented in
	Eqs.~\eqref{eqD1P1},
	\eqref{eqD3P0},
	\eqref{eqD3P1},
	and
	\eqref{eqD3P2}
	are new results.
	For the unequal-mass case,
	relevant for flavored heavy mesons such as the $B_c$ family,
	the corresponding $\mathcal{O}(v^2)$ corrections
	given in
	Eqs.~\eqref{eqBD1P1},
	\eqref{eqBD3P0},
	\eqref{eqBD3P1},
	and
	\eqref{eqBD3P2}
	are also obtained for the first time.
	These results constitute the main contributions of the present work.

	\section{Numerical results and discussion}
\label{sec:Num}

In this section, based on the analytical expressions of the SDCs obtained in the previous section,
we perform a numerical analysis of the $\mathcal O(v^2)$ relativistic effects
in the fragmentation processes of various $P$-wave heavy hadronic states.
The relative values of the LDMEs are given by $\langle 0| \mathcal{O}^{H} |0 \rangle : \langle 0| \mathcal{P}^{H} |0 \rangle = 1 : (2\mu)^2 \langle v^2 \rangle$. $\langle v^2\rangle$ adhere to the velocity power scaling rules and are of the order of $v^2$ as $v^2=\langle v^2\rangle[1+\mathcal{O}(v^4)]$.

\begin{figure}[htbp]
		\centering
		
		\begin{subfigure}{0.48\textwidth}
			\centering
			\includegraphics[width=\linewidth]{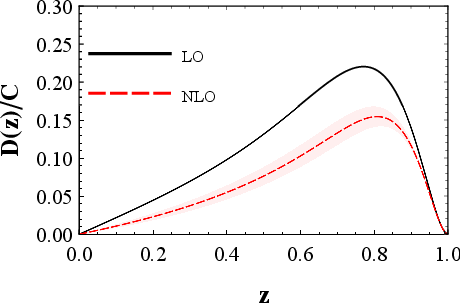}
			\caption{$c \to h_c$}
			\label{fig:c_hc}
		\end{subfigure}
		\hfill
		\begin{subfigure}{0.48\textwidth}
			\centering
			\includegraphics[width=\linewidth]{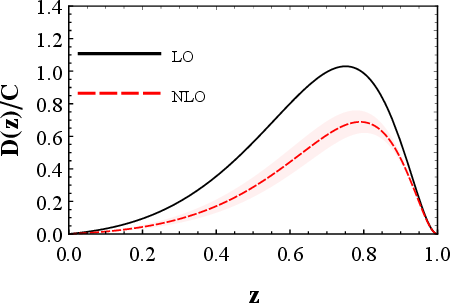}
			\caption{$c \to \chi_{c0}$}
			\label{fig:c_chic0}
		\end{subfigure}
		
		\vspace{0.5cm}
		
		\begin{subfigure}{0.48\textwidth}
			\centering
			\includegraphics[width=\linewidth]{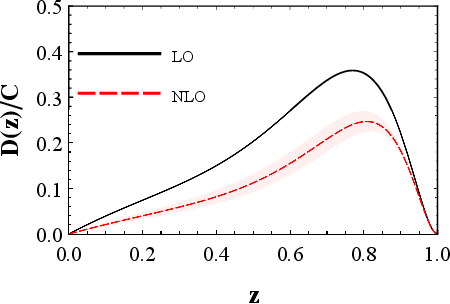}
			\caption{$c \to \chi_{c1}$}
			\label{fig:c_chic1}
		\end{subfigure}
		\hfill
		\begin{subfigure}{0.48\textwidth}
			\centering
			\includegraphics[width=\linewidth]{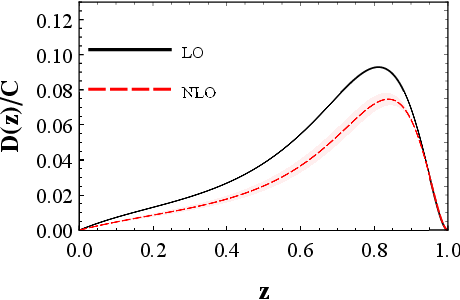}
			\caption{$c \to \chi_{c2}$}
			\label{fig:c_chic2}
		\end{subfigure}
		
		\caption{
			(Color online)
			The heavy-quark fragmentation functions
			$D(c \to h_c)$,
			$D(c \to \chi_{c0})$,
			$D(c \to \chi_{c1})$, and
			$D(c \to \chi_{c2})$
			as functions of the momentum fraction $z$.
			The black solid curves denote the leading-order results, while the red dashed curves include the $\mathcal{O}(v^{2})$ relativistic corrections.
		The shaded bands correspond to the variation of
		$\langle v^{2} \rangle_{c\bar c} = 0.23 \pm 0.05$ in the charmonium system.
			The charm-quark mass is taken as
			$m_c = 1.5~\mathrm{GeV}$.
			The normalization factor is chosen as
			$C = 10^{-2}\,\alpha_s^2 \langle \mathcal{O} \rangle$.
		}
		\label{fig:frag_charmonium}
	\end{figure}

\begin{figure}[htbp]
		\centering
		
		\begin{subfigure}{0.48\textwidth}
			\centering
			\includegraphics[width=\linewidth]{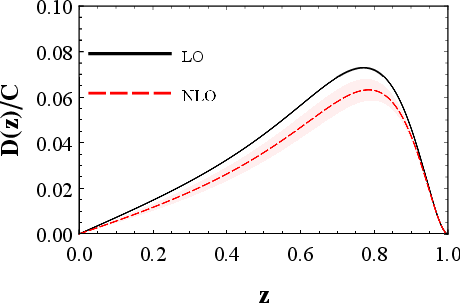}
			\caption{$b \to h_b$}
			\label{fig:b_hb}
		\end{subfigure}
		\hfill
		\begin{subfigure}{0.48\textwidth}
			\centering
			\includegraphics[width=\linewidth]{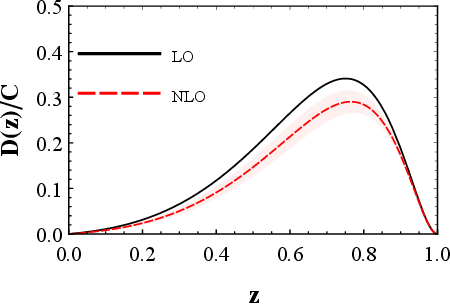}
			\caption{$b \to \chi_{b0}$}
			\label{fig:b_chib0}
		\end{subfigure}
		
		\vspace{0.5cm}
		
		\begin{subfigure}{0.48\textwidth}
			\centering
			\includegraphics[width=\linewidth]{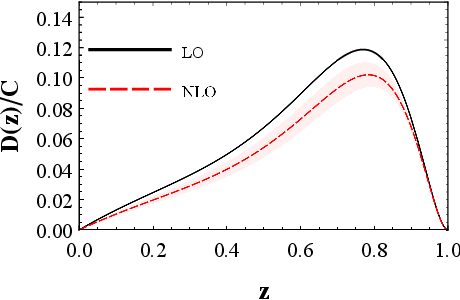}
			\caption{$b \to \chi_{b1}$}
			\label{fig:b_chib1}
		\end{subfigure}
		\hfill
		\begin{subfigure}{0.48\textwidth}
			\centering
			\includegraphics[width=\linewidth]{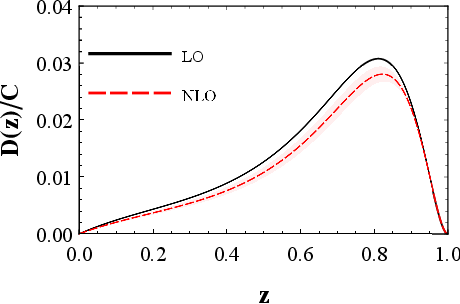}
			\caption{$b \to \chi_{b2}$}
			\label{fig:b_chib2}
		\end{subfigure}
		
		\caption{
			(Color online)
			The heavy-quark fragmentation functions
			$D(b \to h_b)$,
			$D(b \to \chi_{b0})$,
			$D(b \to \chi_{b1})$, and
			$D(b \to \chi_{b2})$
			as functions of the momentum fraction $z$.
			The notation and color coding follow those in
			Figure~\ref{fig:frag_charmonium},
			but with the bottom-quark mass fixed to
			$m_b = 4.7~\mathrm{GeV}$.
		The shaded bands correspond to the variation of
		$\langle v^{2}\rangle_{b\bar b} = 0.10 \pm 0.05$ in the bottomonium system.
			The normalization factor is taken as
			$C = 10^{-4}\,\alpha_s^2 \langle \mathcal{O} \rangle$.
		}
		\label{fig:frag_bottomonium}
	\end{figure}

The charm- and bottom-quark masses are taken from Ref.~\cite{ParticleDataGroup:2018ovx} as
\begin{equation}
	m_c = 1.5~\mathrm{GeV},
	\qquad
	m_b = 4.7~\mathrm{GeV}.
\end{equation}
For the equal-mass (quarkonium) case, the relativistic velocity parameter $v^2$ in the CS channel
is estimated using the Gremm--Kapustin relation~\cite{Gremm:1997dq},
which is given by
\begin{equation}
	v^2
	=
	\frac{M_{Q\bar Q}-2m_Q^{\mathrm{pole}}}
	{m_Q^{\mathrm{QCD}}} ,
\end{equation}
where $m_Q^{\mathrm{pole}}$ denotes the pole mass of the heavy quark,
$m_Q^{\mathrm{QCD}}$ is the quark mass parameter appearing in NRQCD,
and $M_{Q\bar Q}$ represents the experimental mass of the corresponding
heavy quarkonium state.
Here we take  \( m_{Q}^{\mathrm{QCD}}=m_{Q}^{\mathrm{pole}} \) , from which we obtain
\begin{align}
	\begin{aligned}
		v_{c\bar c}^2 = 0.23,	\qquad v_{b\bar b}^2 = 0.1.
	\end{aligned} 
	\label{eq:v_values}
\end{align}
It should be emphasized that the value of $v^2$ extracted from the
Gremm--Kapustin relation
can only be regarded as an order-of-magnitude estimate,
and is subject to sizable theoretical uncertainties.
Following the commonly adopted ranges in the literature,
we take
$\langle v^2\rangle_{c\bar c}$ in the interval $0.23\pm 0.05$
and
$\langle v^2\rangle_{b\bar b}$ in the interval $0.1\pm 0.05$.
Based on these inputs,
we evaluate the integrated ratios of the
$\mathcal O(v^2)$ relativistic corrections
to the LO fragmentation functions in
Table~\ref{tab1}.
In addition,
we further examine the ratios of the
$\mathcal O(v^2)$ relativistic SDCs
to their LO counterparts,
evaluated in the kinematic limits $z \to 0$ and $z \to 1$.
These two limits correspond to the soft and forward
fragmentation regions, respectively.
The corresponding results are summarized in
Table~\ref{tab2},
and compared with those available in
Refs.~\cite{Sang:2009zz,Cui:2025wjq}.
To illustrate the overall size of the
$\mathcal O(v^2)$ relativistic effects more transparently,
we also present the fragmentation functions
of $P$-wave charmonium and bottomonium states
as functions of the momentum fraction $z$.
Specifically,
Figure~\ref{fig:frag_charmonium}
and
Figure~\ref{fig:frag_bottomonium}
show the numerical results for the
charmonium and bottomonium systems, respectively.

For the unequal-mass case,
taking the $B_c$ meson system as a representative example,
we compute the integrated ratios of the
$\mathcal O(v^2)$ relativistic corrections
to the LO fragmentation functions
in the ${}^{1}P_{1}$ and ${}^{3}P_{J}$ ($J=0,1,2$) channels.
The corresponding numerical results are listed in
Table~\ref{tab3},
which allow for a systematic analysis of relativistic effects
in heavy-meson fragmentation processes
with unequal quark masses.


		\begin{table}[htbp]
		\centering
		\caption{
			Ratios of the integrated fragmentation functions. The leading-order fragmentation function is denoted by $D^{(0)}(z)$.
			The fragmentation function from relativistic correction is denoted by $D^{(2)}(z)$.
		For the equal-mass case, $\langle v^{2}\rangle_{c\bar c}=0.23\pm 0.05$ and $\langle v^{2}\rangle_{b\bar b}=0.1\pm 0.05$ are used.
	}
		\label{tab1}
		\renewcommand{\arraystretch}{1.5}  
        \begin{tabular}{l c c c c} 
        \hline \hline 
        & $c \to h_c(^1P_1)$ & $c \to \chi_{c0}(^3P_0)$ & $c \to  \chi_{c1}(^3P_1)$ & $c \to  \chi_{c2}(^3P_2)$ \\
        \hline 
        $\frac{\int dzD^{(2)}(z)}{\int dzD^{(0)}(z)}$ & $-35.36\pm7.69\%$ & $-38.71\pm8.42\%$ & $-36.89\pm8.02\%$ & $-25.73\pm5.59\%$ \\ 
        \hline \hline 
        & $b \to h_b(^1P_1)$ & $b \to \chi_{b0}(^3P_0)$ & $b \to  \chi_{b1}(^3P_1)$ & $b \to  \chi_{b2}(^3P_2)$ \\
        \hline 
        $\frac{\int dzD^{(2)}(z)}{\int dzD^{(0)}(z)}$ & $-15.37\pm7.69\%$ & $-16.83\pm8.42\%$ & $-16.04\pm8.02\%$ & $-11.19\pm5.59\%$ \\ 
        \hline \hline 
        \end{tabular}
	   \end{table}

	\begin{table}[htbp]
	\centering
	\caption{
		Ratios of the $\mathcal{O}(v^2)$ relativistic short-distance coefficients
		to the leading-order ones in the equal-mass case,
		evaluated in the kinematic limits $z \to 0$ and $z \to 1$,
		which correspond to the soft and forward fragmentation regions,
		respectively. These ratios are defined as
		$R_{1,2} = m^2 \left. \left( G^n / F^n \right) \right|_{z \to 0,1}$.
	}
	\label{tab2}
	\renewcommand{\arraystretch}{1.5}  
	\begin{tabular}{@{\extracolsep{1em}} l c c c c @{}}  
		\hline
		\hline
		& 
		$^1P_1$ & $^3P_0$ & $^3P_1$ & $^3P_2$ \\
		\hline
		$R_1$ & $-\frac{23}{10}$  &  $-\frac{23}{10}$ &  $-\frac{21}{10}$ &  $-\frac{17}{10}$ \\ 
		\hline
		$R_2$ & $\frac{7}{50}$  &  $-\frac{17}{1570}$ &  $\frac{7}{85}$ &  $\frac{167}{320}$ \\ 
		\hline
		\hline
	\end{tabular}
\end{table}

	For the equal-mass case, we find that the
	$\mathcal{O}(v^2)$ relativistic corrections are negative
	over most of the momentum-fraction region $z$.
	The relativistic effects are numerically more significant
	in the charmonium system than in bottomonium,
	which can be traced back to the larger relative velocity
	of the heavy quarks in the former.
	This overall trend is consistent with previous observations
	in heavy-quark fragmentation into $S$-wave quarkonia~\cite{Sang:2009zz}.
	
	A comparison among different angular-momentum channels further shows that,
	for CS $P$-wave states,
	the relative size of the $\mathcal{O}(v^2)$ corrections
	is generally larger than that for the corresponding
CS $S$-wave channels.
	This indicates that relativistic effects play a more important role
	in $P$-wave fragmentation processes.
	In addition, the numerical results exhibit a strong dependence
	on the choice of the velocity parameter $v_{Q\bar Q}^{2}$,
	highlighting the sensitivity of the relativistic corrections
	to the internal dynamics of the bound state.
	
	Noticeable differences are also observed among various $P$-wave channels.
	In particular, the relativistic corrections to the ${}^{3}P_{2}$ state
	are significantly smaller than those for the
	${}^{1}P_{1}$, ${}^{3}P_{0}$, and ${}^{3}P_{1}$ channels.
	To further quantify the impact of relativistic effects,
	we evaluate the ratios $R_i$ $(i=1,2)$,
	which measure the contributions of the $\mathcal{O}(v^2)$ corrections
	relative to the LO results.
	These ratios are summarized in Table~\ref{tab2}.
	
	By comparing our results with the corresponding $S$-wave fragmentation
	results in Ref.~\cite{Sang:2009zz,Cui:2025wjq},
	we conclude that, for heavy-quark fragmentation into quarkonium,
	the relativistic corrections in $P$-wave channels
	are generally larger than those in $S$-wave channels.
	This enhancement becomes particularly pronounced
	in the $z \to 1$ forward-fragmentation region,
	where the $\mathcal{O}(v^2)$ corrections have a stronger impact
	on the fragmentation-function distributions.
	Therefore, relativistic corrections at order $v^2$
	are indispensable for precision theoretical predictions
	of $P$-wave heavy-quarkonium production.


	\begin{table}[H]
		\centering
	\caption{
		Ratios of the integrated $\mathcal{O}(v^2)$ relativistic corrections
		to the integrated leading-order fragmentation functions
		for $P$-wave $B_c$ mesons in the
		${}^{1}P_{1}$ and ${}^{3}P_{J}$ ($J=0,1,2$) channels.
	}
		\label{tab3}
		\renewcommand{\arraystretch}{2}  
		\begin{tabular}{@{\extracolsep{1em}} l c c c c @{}}
			\hline
			\hline
			& \multicolumn{4}{c}{$m_1=1.5,\;m_2=4.7$} \\
			\cline{2-5}
			& \(c \to c\bar{b}(^1P_1)\) & \(c \to c\bar{b}(^3P_0)\) & \(c \to  c\bar{b}(^3P_1)\) & \(c \to  c\bar{b}(^3P_2)\) \\
			\hline
			$ \frac{\int dzD^{(2)}(z)}{\langle v^2 \rangle\int dzD^{(0)}(z)}$ & $ -1.54$ & $-1.82$ & $-1.61$ & $-0.44$ \\ 
			\hline
			\hline
			& \multicolumn{4}{c}{$m_1=4.7,\;m_2=1.5$} \\
			\cline{2-5}
			& \(b \to b\bar{c}(^1P_1)\) & \(b \to b\bar{c}(^3P_0)\) & \(b \to  b\bar{c}(^3P_1)\) & \(b \to  b\bar{c}(^3P_2)\) \\
			\hline
			$ \frac{\int dzD^{(2)}(z)}{\langle v^2 \rangle\int dzD^{(0)}(z)}$  & $ -2.97$ & $-2.69$ & $-2.98$ & $-3.01$ \\ 
			\hline
			\hline
		\end{tabular}
	\end{table}
	
For the unequal-mass case, we take the $B_c$ meson as a representative example
and evaluate the ratios of the integrated
$\mathcal{O}(v^2)$ relativistic corrections
to the integrated LO fragmentation functions
for $P$-wave $B_c$ states
in the ${}^{1}P_{1}$ and ${}^{3}P_{J}$ ($J=0,1,2$) channels.
For the fragmentation process $c \to c\bar b$,
the relativistic corrections at order $\mathcal{O}(v^2)$
exhibit a mild dependence on the quantum numbers of the final-state meson.
In particular, the contributions for the
${}^{1}P_{1}$, ${}^{3}P_{0}$, and ${}^{3}P_{1}$ channels
are found to be of comparable magnitude,
while the correction to the ${}^{3}P_{2}$ channel
is noticeably smaller than those of the other $P$-wave states.
In contrast, for the fragmentation process $b \to b\bar c$,
the relativistic corrections show much weaker sensitivity
to the spin and angular-momentum quantum numbers.
The $\mathcal{O}(v^2)$ contributions to the
${}^{1}P_{1}$ and ${}^{3}P_{J}$ ($J=0,1,2$) channels
are numerically similar,
indicating that relativistic effects in this channel
are largely insensitive to the specific $P$-wave configuration.
These results reflect the interplay between relativistic effects
and mass asymmetry in the $B_c$ system,
and highlight qualitative differences between
the $c \to c\bar b$ and $b \to b\bar c$ fragmentation mechanisms.

\section{Application and comparison with full fix-order calculations}
\label{sec:compare}
	
\begin{figure}[tbp]
	\centering %
	\includegraphics[width=.65\textwidth]{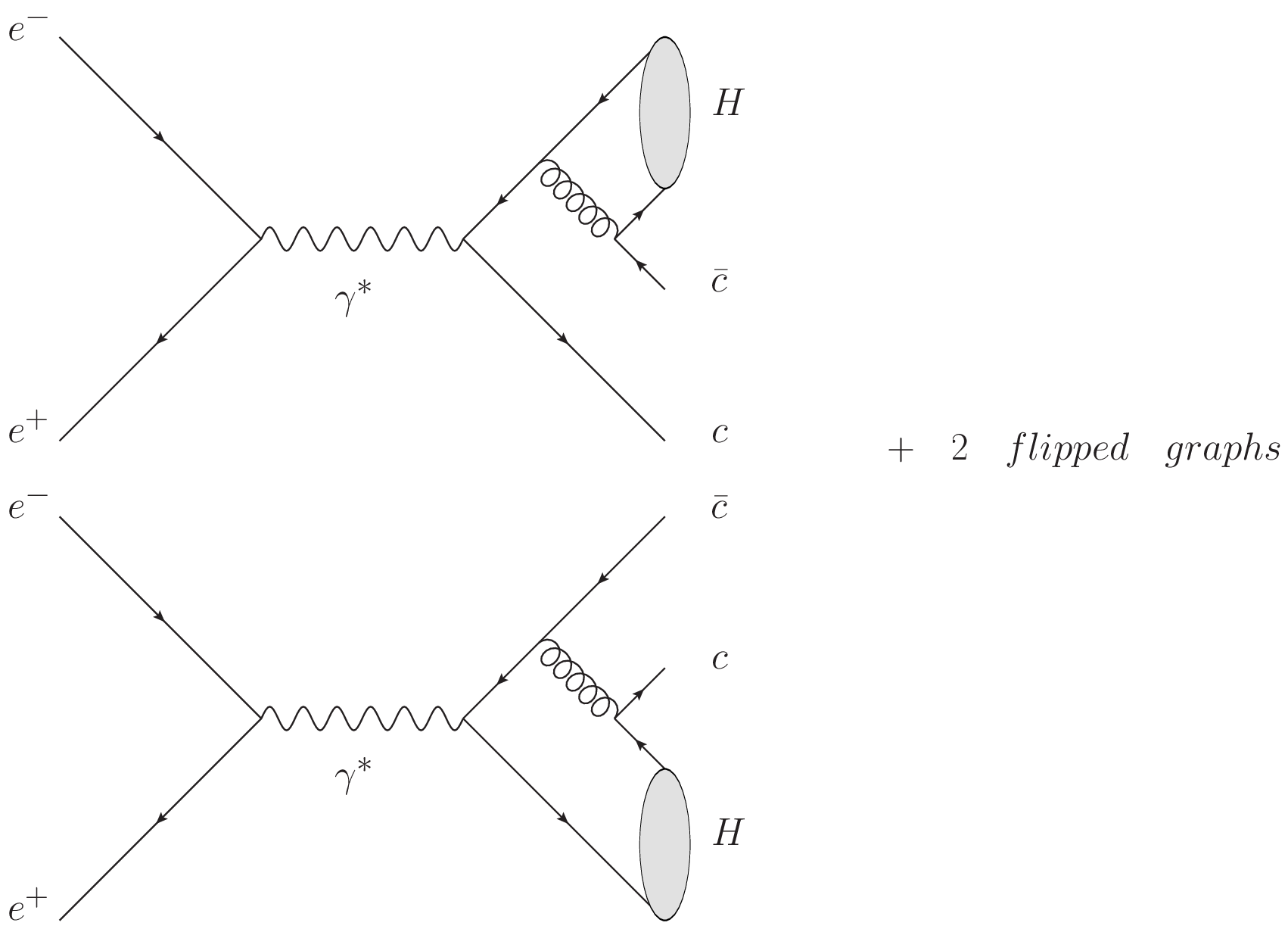}
	\caption{\label{fig:00} Feynman diagrams for $e^+ + e^- \to \gamma^* \to H + X_{c\bar c}$ at quark level.}
\end{figure}

We apply the obtained fragmentation functions to the inclusive production
of $P$-wave charmonium (denoted by $H$) in $e^+e^-$ annihilation via a virtual photon.
The representative Feynman diagrams are shown in Figure~\ref{fig:00}.

Within the fragmentation approximation, the cross sections can be written as
\begin{equation}
	\sigma_{\mathrm{frag}}\!\left(e^+ e^- \to \gamma^* \to H + X_{c\bar c}\right)
	=
	2\,\sigma\!\left(e^+ e^- \to c\bar{c}\right)
	\int_{\delta}^{1}
	D_{c \to H}(z)\, {d}z ,
\end{equation}
where $\sigma(e^+e^- \to c\bar c)$ denotes the charm quark pair production cross section,
$D(z)$ is the fragmentation function for a charm quark fragmenting into a $P$-wave charmonium state,
and $\delta = 4m_c/E_{\mathrm{cm}}$ where $E_{\mathrm{cm}}$ is the center-of-mass energy.

\begin{figure}[htbp]
	\centering
	
		\begin{subfigure}{0.48\textwidth}
		\centering
		\includegraphics[width=\linewidth]{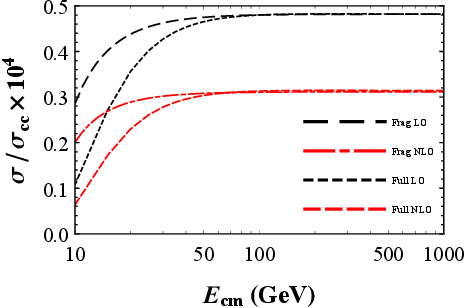}
		\caption{$h_c$ production}
		\label{fig:1P1ff}
	\end{subfigure}
	\hfill
	\begin{subfigure}{0.48\textwidth}
		\centering
		\includegraphics[width=\linewidth]{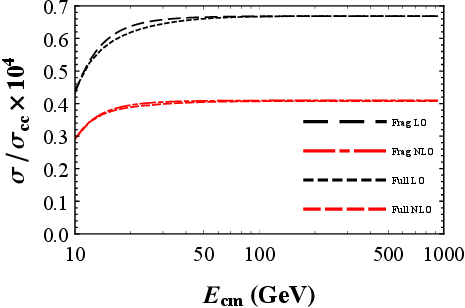}
		\caption{$\chi_{c0}$ production}
		\label{fig:3P0ff}
	\end{subfigure}
	\hfill
	\begin{subfigure}{0.48\textwidth}
		\centering
		\includegraphics[width=\linewidth]{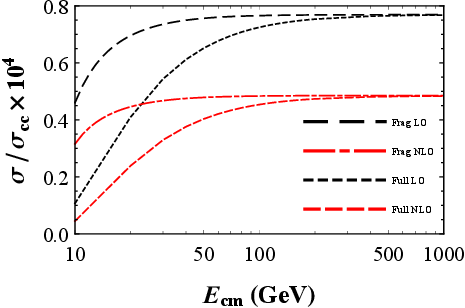}
		\caption{$\chi_{c1}$ production}
		\label{fig:3P1ff}
	\end{subfigure}
	\hfill
	\begin{subfigure}{0.48\textwidth}
		\centering
		\includegraphics[width=\linewidth]{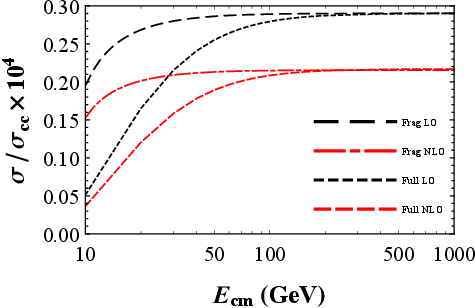}
		\caption{$\chi_{c2}$ production}
		\label{fig:3P2ff}
	\end{subfigure}
	\caption{
		(Color online)
		Ratios of cross sections $\sigma(e^+e^- \to H + X_{c\bar c})$ and $\sigma(e^+e^- \to c\bar c)$ as a function of the center-of-mass energy $E_{\mathrm{cm}}$. Here, ``Frag LO'' and ``Frag NLO'' denote the leading-order and next-to-leading-order results in the fragmentation approximation, while ``Full LO'' and ``Full NLO'' denote the leading-order and next-to-leading-order results from the full fixed-order calculation, respectively.
	}
	\label{fig:ff}
\end{figure}

Based on the above fragmentation expression,
we further plot the ratios between the scattering cross sections of the charmonium production and those of the charm quark pair production as functions of $E_{\mathrm{cm}}$ as shown in Figure~\ref{fig:ff}.
The corresponding ratios of full fixed-order perturbative QCD calculations at LO \cite{Liu:2003jj} and up to $\mathcal{O}(v^2)$ are also plotted in this figure for comparison\footnote{In our other recent work, we revisit the relativistic corrections to the inclusive processes $e^+e^- \to J/\psi + X_{c\bar c}$ and $e^+e^- \to J/\psi + X_{\mathrm{non}\,c\bar c}$ at $B$-factory energies. We consider the expansions of all the kinematic final-state momenta by relative momentum $\mathbf q$. The corresponding paper is in preparation\cite{Jiang:2026rr}.}.

From the Figure~\ref{fig:ff}, one can directly observe that,
similar to the LO case \cite{Liu:2003jj},
the fragmentation and full calculations also remain in good agreement in the high-energy limit for the NLO$(v^2)$ case.
These results provide a twofold validation of both the fragmentation functions and the full NLO$(v^2)$ calculations to a certain extent.
In comparison, there are clear deviations persist at low energies.
This implies that the fragmentation function calculations lack accuracy in the low-energy region, where the full NLO calculations need to be employed instead.


\section{Summary}
\label{sec:summary}

	In this work, within the framework of NRQCD factorization,
	we have systematically investigated the
	$\mathcal O(v^2)$ relativistic corrections
	to heavy-quark fragmentation into CS $P$-wave heavy hadrons,
	using the Collins--Soper operator definition of fragmentation functions
	in the axial gauge.
	Taking the known LO results in
	Refs.~\cite{Yuan:1994hn,Ma:1995vi} as benchmarks,
	we reproduce and verify the corresponding LO SDCs.
Building upon these benchmarks,
we derive analytic expressions for the
	$\mathcal O(v^2)$ relativistic corrections
	to heavy-quark fragmentation into
the spin-singlet $\,{}^{1}P_{1}\,(h_Q)\,$
	and spin-triplet
	$\,{}^{3}P_{J}\,(\chi_{QJ},\,J=0,1,2)\,$
	states,
	covering both equal-mass quarkonium systems
	and unequal-mass configurations.
	The results presented in this paper
	provide a systematic extension of the existing fragmentation-function framework
	for $P$-wave heavy hadrons
	and can be directly applied to precision theoretical predictions
	for the production of
	$h_Q$ and $\chi_{QJ}$ states
	at high-energy colliders.
	We also perform a detailed comparison between the fragmentation-approximation results
	and the full fixed-order calculation, and find that they
become consistent
	in the high-energy region.
	Moreover,
	the present analysis establishes a unified theoretical basis
	for future studies of higher-order relativistic corrections
	in heavy-quark fragmentation processes.

	\section*{Acknowledgements}

	This work was supported by the National Natural Science Foundation of China (No. 11705078, 12575087).

	\appendix
	\section*{Appendix}
	\label{app:Bc_fragmentation}
	\setcounter{equation}{0}  
	\renewcommand{\theequation}{A.\arabic{equation}}  

	\subsection*{A. Order-by-order expansion of the squared amplitude for $P$-wave states}
	\label{app:amplitude_P}
	For definiteness, we illustrate the procedure in the equal-mass limit
	$m_1=m_2=m$ (quarkonium), where $E(\mathbf q)=\sqrt{m^2+\mathbf q^2}$ and
	$E(\mathbf q)\to m$ as $\mathbf q\to \mathbf 0$.
	To organize the velocity expansion, we rewrite the invariant amplitude
	in Eq.~\eqref{AP} as
	\begin{equation}
		\mathcal{M}(q) \equiv \widetilde{\mathcal{M}}\!\left(q,\, E(\mathbf{q})\right),
	\end{equation}
	treating $\widetilde{\mathcal{M}}(q,E)$ as a function of two independent variables
	$q$ and $E$.
	In the limit $q\to 0$, one has $E(\mathbf{q})\to m$.
	For the $P$-wave channel, $\mathcal{M}(q)$ is odd in $q$ and hence
	$\widetilde{\mathcal{M}}(0,m)=0$.
	
	The relativistic dispersion relation is expanded as
	\begin{equation}
		E(\mathbf{q}) = m + \frac{\mathbf{q}^2}{2m}
		+ \mathcal O\!\left(\frac{\mathbf{q}^4}{m^3}\right).
	\end{equation}

Following Eq.~\eqref{MP}, the $P$-wave component of the squared amplitude,
with trace evaluations and other intermediate steps suppressed,
can be expanded up to $\mathcal O(v^{2})$ as
\begin{align}
	|\mathcal M_P|^2
	&=
	\mathbf q^{\,2}
	\Bigg\{
	\epsilon^{\alpha}(L_z)\,
	\epsilon^{*\alpha'}(L_z)
	\left[
	\frac{\partial \widetilde{\mathcal M}}{\partial q^\alpha}(0,m)\,
	\frac{\partial \widetilde{\mathcal M}^{*}}{\partial q^{\alpha'}}(0,m)
	\right]
	\Bigg\}
	\nonumber\\
	&\quad
	-\frac{\mathbf q^{\,4}}{30}
	\Bigg\{
	\epsilon^{\alpha}(L_z)\,
	\epsilon^{*\alpha'}(L_z)\,
	I_{\alpha'}^{\beta\gamma\delta}
	\left[
	\frac{\partial \widetilde{\mathcal M}}{\partial q^\alpha}(0,m)\,
	\frac{\partial^{3}\widetilde{\mathcal M}^{*}}
	{\partial q^\beta\,\partial q^\gamma\,\partial q^\delta}(0,m)
	+\mathrm{H.c.}
	\right]
	\Bigg\}
	\nonumber\\
	&\quad
	+
	\frac{\mathbf q^{\,4}}{2m}\,
	\frac{\partial}{\partial E}
	\Bigg\{
	\epsilon^{\alpha}(L_z)\,
	\epsilon^{*\alpha'}(L_z)
	\left[
	\frac{\partial \widetilde{\mathcal M}}{\partial q^\alpha}(0,E)\,
	\frac{\partial \widetilde{\mathcal M}^{*}}{\partial q^{\alpha'}}(0,E)
	\right]
	\Bigg\}
	+\mathcal O(v^{4}) .
	\label{eq:Pwave_M2_derivative}
\end{align}

For the spin-triplet ${}^{3}P_J$ states ($J=0,1,2$), the physical polarization
degrees of freedom arise from the coupling of the orbital and spin
polarizations.
By combining the orbital angular momentum $L=1$ and the spin $S=1$,
the total angular-momentum polarization tensors can be defined as
\begin{equation}
	\sum_{L_z,S_z}
	\epsilon^{\alpha}(L_z)\,
	\epsilon^{\mu}(S_z)\,
	\langle 1 L_z;\, 1 S_z \mid J J_z \rangle
	=
	\epsilon_{(J)}^{\alpha\mu}(J_z),
	\label{eq:Pwave_pol_coupling}
\end{equation}
where $\langle 1 L_z;\, 1 S_z \mid J J_z \rangle$ denotes the Clebsch--Gordan
coefficient for coupling orbital angular momentum $L=1$ and spin $S=1$
to form a state with total angular momentum $J$.
The polarization sum relations for the ${}^{3}P_J$ states are given by
\begin{align}
	\sum_{J_z}
	\epsilon_{(0)}^{\alpha\mu}(J_z)\,
	\epsilon_{(0)}^{*\alpha'\mu'}(J_z)
	&=
	\frac{1}{d-1}\,
	I^{\alpha\mu}\,
	I^{\alpha'\mu'},
	\label{eq:d_dim_spin0_sum}
	\\
	\sum_{J_z}
	\epsilon_{(1)}^{\alpha\mu}(J_z)\,
	\epsilon_{(1)}^{*\alpha'\mu'}(J_z)
	&=
	\frac{1}{2}
	\left(
	I^{\alpha\alpha'}\, I^{\mu\mu'}
	-
	I^{\alpha\mu'}\, I^{\alpha'\mu}
	\right),
	\label{eq:d_dim_spin1_sum}
	\\
	\sum_{J_z}
	\epsilon_{(2)}^{\alpha\mu}(J_z)\,
	\epsilon_{(2)}^{*\alpha'\mu'}(J_z)
	&=
	\frac{1}{2}
	\left(
	I^{\alpha\alpha'}\, I^{\mu\mu'}
	+
	I^{\alpha\mu'}\, I^{\alpha'\mu}
	\right)
	-
	\frac{1}{d-1}\,
	I^{\alpha\mu}\,
	I^{\alpha'\mu'} .
	\label{eq:d_dim_spin2_sum}
\end{align}

		\setcounter{equation}{0}
	\renewcommand{\theequation}{B.\arabic{equation}}

\subsection*{B. Short-distance coefficients for $P$-wave states in the equal-mass case}

	\label{B}

	For the $^1P_1$, we have
\begin{align}
	&F^{^1P_1}(z)
	=\frac{16 \alpha_s^2z (z-1)^2 }{243 m^5 (z-2)^8} \nonumber\\
	&\times(9 z^6-56 z^5+140 z^4-160 z^3+176 z^2-128 z+64)	,
	\\[4pt]
	&G^{^1P_1}(z)
	=-
	\frac{8\alpha_s^2\,z(1-z)^2}{18225m^7(2-z)^{10}}\nonumber\\
	&\times(3675 z^8-39844 z^7+169032 z^6-373552 z^5+483904 z^4-535360 z^3+498560 z^2-295680 z+88320),
	\label{eqD1P1}
\end{align}

For the $^3P_0$, we have
\begin{align}
	&F^{^3P_0}(z)
	=
	\frac{16\alpha_s^2z (z-1)^2  }{729 m^5 (z-2)^8}\nonumber\\
	&\times(59 z^6-376 z^5+1060 z^4-1376 z^3+528 z^2+384 z+192),
	\\[4pt]
	&G^{^3P_0}(z)
	=-
	\frac{8 \alpha_s^2\,z(1-z)^2}{18225m^7(2-z)^{10}}\nonumber\\
	&\times(7035 z^8-80404 z^7+375432 z^6-967792 z^5+1397824 z^4-961600 z^3+22400 z^2+119040 z+88320),
	\label{eqD3P0}
\end{align}

For the $^3P_1$, we have
\begin{align}
	&F^{^3P_1}(z)
	=
	\frac{64 \alpha_s^2  z(z-1)^2}{729 m^5 (z-2)^8}\nonumber\\
	&\times(7 z^6-54 z^5+202 z^4-408 z^3+496 z^2-288 z+96),
	\\[4pt]
	&G^{^3P_1}(z)
	=-
	\frac{64\alpha_s^2\,z(1-z)^2}{18225m^7(2-z)^{10}}\nonumber\\
	&\times(515 z^8-6393 z^7+33789 z^6-103684 z^5+200068 z^4-254240 z^3+196080 z^2-86400 z+20160),
	\label{eqD3P1}
\end{align}

For the $^3P_2$, we have
\begin{align}
	&F^{^3P_2}(z)
	=
	\frac{128\alpha_s^2  z(z-1)^2 }{3645 m^5 (z-2)^8}\nonumber\\
	&\times(23 z^6-184 z^5+541 z^4-668 z^3+480 z^2-192 z+48),
	\\[4pt]
	&G^{^3P_2}(z)
	=-
	\frac{32 \alpha_s^2\,z(1-z)^2}{18225m^7(2-z)^{10}}\nonumber\\
	&\times(1593 z^8-19924 z^7+99054 z^6-249760 z^5+337240 z^4-267424 z^3+130976 z^2-38784 z+6528).
	\label{eqD3P2}
\end{align}
	
	\setcounter{equation}{0}
	\renewcommand{\theequation}{C.\arabic{equation}}

	\subsection*{C. Short-distance coefficients for $P$-wave states in the unequal-mass case}
	
	For the $^1P_1$, we have
\begin{align}
	&	F^{^1P_1}(z)
	=-\frac{2 \alpha_s^2 z(z-1)^2}{729 M^5 ({y_1}-1)^3 {y_1}^2 ({y_1} z-1)^8}\nonumber\\
	&\times\big[ \left(46 y_1^6-78 y_1^5+41 y_1^4\right) z^6\nonumber \\
	&+\left(24 y_1^6-240 y_1^5+318 y_1^4-144 y_1^3\right) z^5\nonumber
	\\
	&+\left(32 y_1^6-128 y_1^5+514 y_1^4-544 y_1^3+210 y_1^2\right) z^4\nonumber
	\\
	&+\left(32 y_1^5-224 y_1^4+96 y_1^3+56 y_1^2-56 y_1\right) z^3\nonumber
	\\
	&+\left(32 y_1^4-32 y_1^3+154 y_1^2-130 y_1+45\right) z^2\nonumber
	\\
	&+\left(-24 y_1^2-6\right) z\nonumber
	\\
	&+6 \big] ,
	\\[4pt]
	&	G^{^1P_1}(z)
	=
	\frac{\alpha_s^2\,z(1-z)^2}{36450 M^7 (y_1-1)^5 {y_1}^4 ({y_1} z-1)^{10}} \nonumber
	\\
	&\times\big[ \left(-2300 y_1^{11}+19750 y_1^{10}-50040 y_1^9+62025 y_1^8-37780 y_1^7+9395 y_1^6\right) z^8\nonumber
	\\
	&+\left(-1200 y_1^{11}+25120 y_1^{10}-142440 y_1^9+309154 y_1^8-355822 y_1^7+209282 y_1^6-51174 y_1^5\right) z^7\nonumber
	\\
	&+\left(-1600 y_1^{11}+20160 y_1^{10}-121200 y_1^9+443628 y_1^8-799484 y_1^7+834909 y_1^6-465468 y_1^5+110265 y_1^4\right) z^6\nonumber
	\\
	&+\left(1600 y_1^{10}-6480 y_1^9+36112 y_1^8-283976 y_1^7+614186 y_1^6-725772 y_1^5+434490 y_1^4-107540 y_1^3\right) z^5\nonumber
	\\
	&+\left(-22464 y_1^8+130792 y_1^7-158992 y_1^6+81244 y_1^5+87965 y_1^4-114220 y_1^3+38725 y_1^2\right) z^4\nonumber
	\\
	&+\left(1600 y_1^8-13680 y_1^7+50560 y_1^6-200120 y_1^5+277350 y_1^4-236830 y_1^3+112070 y_1^2-24550 y_1\right) z^3\nonumber
	\\
	&+\left(-1600 y_1^7+960 y_1^6+13360 y_1^5+3460 y_1^4-5540 y_1^3+12315 y_1^2-8900 y_1+3375\right) z^2\nonumber
	\\
	&+\left(1200 y_1^5-3120 y_1^4-5400 y_1^3+4110 y_1^2-1800 y_1-450\right) z\nonumber
	\\
	&-300 y_1^3+1530 y_1^2-900 y_1+450 \big],
	\label{eqBD1P1}
\end{align}

For the $^3P_0$, we have
\begin{align}
	&	F^{^3P_0}(z)
	=
	-	\frac{2 \alpha_s^2 (z-1)^2 z}{729 M^5 ({y_1}-1)^3 {y_1}^2 ({y_1} z-1)^8}\nonumber\\
	&\times\big[ \left(96 y_1^8-144 y_1^7-10 y_1^6+34 y_1^5+27 y_1^4\right) z^6\nonumber
	\\
	&+\left(-480 y_1^7+888 y_1^6-248 y_1^5-130 y_1^4-48 y_1^3\right) z^5\nonumber
	\\
	&+\left(832 y_1^6-1776 y_1^5+1074 y_1^4-232 y_1^3+150 y_1^2\right) z^4\nonumber
	\\
	&+\left(-800 y_1^5+1728 y_1^4-1328 y_1^3+496 y_1^2-168 y_1\right) z^3\nonumber
	\\
	&+\left(832 y_1^4-1872 y_1^3+1682 y_1^2-714 y_1+135\right) z^2\nonumber
	\\
	&+\left(-480 y_1^3+936 y_1^2-648 y_1+162\right) z\nonumber
	\\
	&+96 y_1^2-144 y_1+54 \big] ,
	\\[4pt]
	&	G^{^3P_0}(z)
	=-
	\frac{ \alpha_s^2 (z-1)^2 z}{36450 M^7 ({y_1}-1)^5 {y_1}^4 ({y_1} z-1)^{10}}\nonumber\\
	&\times\big[\left(4800 y_1^{13}-29760 y_1^{12}+56380 y_1^{11}-38910 y_1^{10}+2940 y_1^9+2705 y_1^8+6560 y_1^7-4985 y_1^6\right) z^8\nonumber
	\\
	&+\left(-33600 y_1^{12}+210960 y_1^{11}-448080 y_1^{10}+418800 y_1^9-154154 y_1^8-3278 y_1^7-6842 y_1^6+18354 y_1^5\right) z^7\nonumber
	\\
	&+\left(94400 y_1^{11}-572800 y_1^{10}+1301120 y_1^9-1515348 y_1^8+984384 y_1^7-352879 y_1^6+98648 y_1^5-45355 y_1^4\right) z^6\nonumber
	\\
	&+\left(-147200 y_1^{10}+844720 y_1^9-1930912 y_1^8+2532736 y_1^7-2122226 y_1^6+1125452 y_1^5-379970 y_1^4+94140 y_1^3\right) z^5\nonumber
	\\
	&+\left(163200 y_1^9-985536 y_1^8+2321288 y_1^7-3199208 y_1^6+2853776 y_1^5-1613175 y_1^4+540880 y_1^3-104175 y_1^2\right) z^4\nonumber
	\\
	&+\left(-147200 y_1^8+991120 y_1^7-2469200 y_1^6+3470000 y_1^5-3032510 y_1^4+1633810 y_1^3-499150 y_1^2+73650 y_1\right) z^3\nonumber
	\\
	&+\left(94400 y_1^7-592000 y_1^6+1336000 y_1^5-1653820 y_1^4+1232560 y_1^3-541625 y_1^2+123000 y_1-10125\right) z^2\nonumber
	\\
	&+\left(-33600 y_1^6+174960 y_1^5-325440 y_1^4+334560 y_1^3-210150 y_1^2+75600 y_1-12150\right) z\nonumber
	\\
	&+4800 y_1^5-20160 y_1^4+29820 y_1^3-26250 y_1^2+15300 y_1-4050  \big] ,
	\label{eqBD3P0}
\end{align}

For the $^3P_1$, we have
\begin{align}
	&	F^{^3P_1}(z)
	=
	-	\frac{4 \alpha_s^2 (z-1)^2 z}{729 M^5 ({y_1}-1)^3 {y_1}^2 ({y_1} z-1)^8}\nonumber\\
	&\times\big[ \left(12 y_1^6-26 y_1^5+17 y_1^4\right) z^6\nonumber\\
	&+\left(-16 y_1^6+46 y_1^4-48 y_1^3\right) z^5\nonumber\\
	&+\left(16 y_1^6-32 y_1^5+94 y_1^4-120 y_1^3+90 y_1^2\right) z^4\nonumber\\
	&+\left(16 y_1^5-80 y_1^4+40 y_1^3+8 y_1^2-56 y_1\right) z^3\nonumber\\
	&+\left(16 y_1^4-64 y_1^3+160 y_1^2-94 y_1+45\right) z^2\nonumber\\
	&+(-24 y_1-6) z+6 \big] ,
	\\[4pt]
	&	G^{^3P_1}(z)
	=
	\frac{ \alpha_s^2 (z-1)^2 z}{18225  M^7 ({y_1}-1)^5 {y_1}^4 ({y_1} z-1)^{10}}\nonumber\\
	&\times\big[ \left(-600 y_1^{11}+5440 y_1^{10}-15040 y_1^9+19955 y_1^8-13020 y_1^7+3595 y_1^6\right) z^8\nonumber\\
	&+\left(800 y_1^{11}-4480 y_1^{10}-4460 y_1^9+39562 y_1^8-71946 y_1^7+55266 y_1^6-17382 y_1^5\right) z^7\nonumber\\
	&+\left(-800 y_1^{11}+5680 y_1^{10}-15060 y_1^9+54914 y_1^8-122372 y_1^7+169427 y_1^6-121124 y_1^5+38905 y_1^4\right) z^6\nonumber\\
	&+\left(800 y_1^{10}-240 y_1^9-4664 y_1^8-48728 y_1^7+141698 y_1^6-209556 y_1^5+150650 y_1^4-50420 y_1^3\right) z^5\nonumber\\
	&+\left(-7232 y_1^8+25156 y_1^7+18414 y_1^6-75068 y_1^5+124755 y_1^4-93700 y_1^3+35725 y_1^2\right) z^4\nonumber\\
	&+\left(800 y_1^8-13840 y_1^7+76720 y_1^6-221100 y_1^5+275870 y_1^4-212970 y_1^3+93990 y_1^2-24550 y_1\right) z^3\nonumber\\
	&+\left(-800 y_1^7+4080 y_1^6-14060 y_1^5+42670 y_1^4-35580 y_1^3+21605 y_1^2-7100 y_1+3375\right) z^2\nonumber\\
	&+\left(1800 y_1^4-9840 y_1^3+6870 y_1^2-3000 y_1-450\right) z\nonumber\\
	&-300 y_1^3+1410 y_1^2-900 y_1+450 \big] ,
	\label{eqBD3P1}
\end{align}

For the $^3P_2$, we have
\begin{align}
	&	F^{^3P_2}(z)
	=-
	\frac{8 \alpha_s^2 (z-1)^2 z}{3645 M^5 ({y_1}-1)^3  ({y_1} z-1)^8}\nonumber\\
	&\times\big[ \left(12 y_1^6-36 y_1^5+91 y_1^4-112 y_1^3+60 y_1^2\right) z^6\nonumber\\
	&+\left(-24 y_1^5+132 y_1^4-412 y_1^3+472 y_1^2-240 y_1\right) z^5\nonumber\\
	&+\left(92 y_1^4-336 y_1^3+702 y_1^2-608 y_1+300\right) z^4\nonumber\\
	&+\left(-40 y_1^3-96 y_1^2+188 y_1-232\right) z^3\nonumber\\
	&+\left(92 y_1^2-108 y_1+151\right) z^2\nonumber\\
	&+(-24 y_1-36) z\nonumber\\
	&+12 \big] ,
	\\[4pt]
	&	G^{^3P_2}(z)
	=
	\frac{2 \alpha_s^2 (z-1)^2 z}{18225 M^7 ({y_1}-1)^5 {y_1}^2 ({y_1} z-1)^{10}}\nonumber\\
	&\times\big[ \left(-120 y_1^{11}+1068 y_1^{10}-3970 y_1^9+11025 y_1^8-19290 y_1^7+20179 y_1^6-11432 y_1^5+2900 y_1^4\right) z^8\nonumber\\
	&+\left(480 y_1^{10}-4296 y_1^9+20376 y_1^8-64134 y_1^7+113144 y_1^6-117034 y_1^5+65912 y_1^4-16896 y_1^3\right) z^7\nonumber\\
	&+\left(-1520 y_1^9+14224 y_1^8-59678 y_1^7+158799 y_1^6-248022 y_1^5+242893 y_1^4-134960 y_1^3+35680 y_1^2\right) z^6\nonumber\\
	&+\left(2480 y_1^8-18832 y_1^7+52804 y_1^6-107980 y_1^5+141836 y_1^4-143588 y_1^3+88592 y_1^2-28560 y_1\right) z^5\nonumber\\
	&+\left(-2640 y_1^7+14016 y_1^6-6038 y_1^5-29101 y_1^4+81466 y_1^3-69027 y_1^2+25304 y_1+1500\right) z^4\nonumber\\
	&+\left(2480 y_1^6-15952 y_1^5+27632 y_1^4-45830 y_1^3+35408 y_1^2-14818 y_1-1160\right) z^3\nonumber\\
	&+\left(-1520 y_1^5+6304 y_1^4+494 y_1^3-2783 y_1^2+3158 y_1+755\right) z^2\nonumber\\
	&+\left(480 y_1^4-2376 y_1^3+756 y_1^2-696 y_1-180\right) z\nonumber\\
	&-120 y_1^3+468 y_1^2-120 y_1+60 \big] .
	\label{eqBD3P2}
\end{align}

\end{document}